# The collaboration of Mileva Marić and Albert Einstein

Estelle Asmodelle

*University of Central Lancashire*

School of Computing, Engineering and Physical Sciences,

Preston, Lancashire, UK PR1 2HE.

e-mail: e.asmodelle@uq.net.au; Phone: +61 418 676 586

---

This is a contemporary review of the involvement of Mileva Marić, Albert Einstein's first wife, in his theoretical work between the period of 1900 to 1905. Separate biographies are outlined for both Mileva and Einstein, prior to their attendance at the Swiss Federal Polytechnic in Zürich in 1896. Then, a combined journal is described, detailing significant events. In addition to a biographical sketch, comments by various authors are compared and contrasted concerning two narratives: first, the sequence of events that happened and the couple's relationship at particular times econd, the contents of letters from both Einstein and Mileva. Some interpretations of the usage of pronouns in those letters during 1899 and 1905 are re-examined, and a different hypothesis regarding the usage of those pronouns is introduced. Various papers are examined and the content of each subsequent paper is compared to the work that Mileva was performing. With a different take, this treatment further suggests that the couple continued to work together much longer than other authors have indicated. We also evaluate critics and supporters of the hypothesis that Mileva was involved in Einstein's work, and refocus this within a historical context, in terms of women in science in the late 19th century. Finally, the definition of collaboration(co-authorship, specifically) is outlined. As a result, recommendations are stated, the first of which is that Mileva should be seriously considered as an honorary co-author of one, possibly two, papers, and second of which is that a serious inquiry should be made concerning the extent of Mileva Marić's involvement in Albert Einstein's published works between 1902 and 1905.

---

## 1 Introduction

Albert Einstein is the most famous physicist to date, and his name is synonymous with genius. The work by Einstein during his life between 1901 and 1921 is considered the most valuable contribution to physics by one person, surpassing even that of Issac Newton. While some priority disputes, over the originality of Einstein's work, have lingered, most are contentious and lacking any real credibility. Einstein's various papers, including Brownian motion, the photoelectric effect, the special theory of rlativity, and the general theory of relativity, remain foundations of modern physics. This article, incidently, was completed on Einstein's birthday, the 14th March 2015, 136 years after his birth. Our understanding of Einstein's life seems solidified in history, for most believe all that is to be known of his life has been well documented.

However, there has been some debate since the 1980s as to the involvement of Einstein's first wife, Mileva Marić, in his work between 1900 and 1905. The impetus for this controversy was the release, by the family, of old letters between Einstein and Mileva. These were later published in the book entitled *Albert Einstein/Mileva Maric: the love letters* [1]. Many letters can be found, amidst other documents of the period, in an anthology entitled *The collected papers of Albert Einstein: the Early Years, 1879-1902, Volume 1*, edited by John Stachel, David C. Cassidy, and Robert Schulmann [2]. This work was published in 1987 as the first volume of the Einstein Papers Project. Both these books are considered the most valuable primary reference material for Einstein's early life. However, many of the letters Mileva wrote to Einstein are missing or destroyed, as well as Einstein's original drafts of his old papers [3]. The contents of the existing letters and the absence of most of Mileva's letters started a silent controversy that has continued since the mid-1980s.

Most of the debate seems to be divided between two schools of thoughthe first being that Mileva had no involvement and only acted as a sounding board, and the second being that Mileva was a genius who did most of the mathematical treatments and was co-author of all the work during this period but has never received any credit. The conservative view is that Mileva was merely a loving wife who supported Einstein and was simply an intelligent sounding board for his ideas. This is primarily the accepted viewpoint by most historians today. This debate remained relatively academic and was not known by the general community until the release of a PBS documentary entitled *Einstein's wife —the life of Mileva Marić Einstein* [4]. Since the release of the documentary, many books have surfaced which have taken the debate to a new level.

In this review, we will look at: some of the remarks found in letters, including the historical context in which they were written; athe sequence of events leading up to Einstein's 1905 *annus mirabilis* (or miracle year); andEinstein and Mileva's divorce. We will also review Einstein's and Mileva's academic achievements and failures. In order to perform such an analysis, we must first examine the social perception of women scientists during the period in question. For the emancipation of women in academic institutions during late 19th century, Europe, is as relevant to this review as is the relationship between Mileva and Einstein as described by the letters of the time.

**2 Educational Opportunities for Women in the 19th Century**

Women during the 19th century were defined by their marriage to a man. In fact, women were not permitted to vote in most of Europe during this century; the first occurence of women's suffrage in Europe was in Finland in 1906, andthe last was in Liechtenstein in 1984 [5]. The 19th century was very much a time of change for women's rights, but it would be a very slow change indeed. Essentially, women had no civil rights and suffered institutionalised sexism, whereby administrators doubted their intellectual capacities and their right to participate in university education [6].

*Gentlemen of science*

In Sue Rosser's book *Women, science, and myth*, the author depicts the attitude of the time: "Science had been a domain open to the privileged gentlemen of science." By the end of the 19th century, men still dominated the sciences but leading up to the end of that century, women participated increasingly in the scientific culture that had become part of everyday life. The women who did pursue education were exceptional pioneers in the expanding fields of science, and many women, still undiscovered, played crucial roles as educators, observers, and explorers of scientific knowledge [6].

*The new world*

The significant changes that occurred in the late 19th century, characterized by the transition from the autocratic rule of monarchs to the bourgeois republic, removed many of the feudal constraints on an industrial revolution. Life in the 19th century was changing dramatically, and women wanted to be part of the new world.

Elizabeth Garrett Anderson was the first female in Britain to gain a medical qualification in 1865. Institutions in Europe were almost on par with Britain. Switzerland Universities began to admit female students after 1867, but women at that time were not permitted to matriculate at either German or Austrian universities until the turn of the twentieth century [7].

*Gasthörer*

Matthias Tomczak, in a paper entitled "Mileva Maric, an unfulfilled career in science," points out that there is confusion about the enrolment status of women between 1867 and 1917 at the Swiss Federal Polytechnic in Zürich. The official student statistic of the institution does not show any enrolled women before 1917. In view of the official student statistics, this has to be understood to mean that before 1917, women were enlisted as *Gasthörer* (or auditors) within University life. This meant that much of a woman's student life during this period was without academic credit [8]. The transition from *Gasthöre* to women attaining university credit would happen slowly. In the late 19th century the Swiss Federal Polytechnic in Zürich proclaimed itself an institution in which women could graduate.

There were a few early exceptions, however. In 1874, Sofia Kovalesky was the first women in modern Europe to gain a Ph.D. at the University of Göttingen. Her research involved applying the theory of differential equations to the study of the shape of the rings of Saturn [6]. In the United States, the first doctoral degree was issued a few years later, in 1886, to Winifred Edgerton Merrill in the field of mathematics from Columbia University [9].

*Women in science*

In the late 19th century in Europe, women were starting to enter science, in the fields of physics and mathematics. However, they suffered discrimination, and only the strong-minded women could deal with the daily sexism [7]. Such discrimination and sexism would eventually change, but it would take a century to redefine women in science, not only in Europe but across the world.

Many women in science in the 21st century are still treated with less respect than their male peers [10]. In fact, the perception of women in physics is still not encouraging. At most, only 18% of female Ph.Ds. are graduates in physics within the US, compared with 49.5% in neuroscience [11]. It is hard to imagine, then, the difficulties that women experienced during Mileva's time. the young, shy Mileva Marić had to face as she began to pursue a career in science.

## 3 Early Biography—1875 to 1896: Mileva Marić

Mileva Marić was born on the 19th December 1875 in Titel, Vojvodina, within the then Austro-Hungarian Empire (what is now Serbia). Mileva was the eldest of three children of Marija and Miloš Marić. Her family was quite wealthy, for Miloš was the Confidential Supervisor at the Royal Court of Justice [3]. The family was of Eastern Orthodox Christian faith. As a young girl, Mileva was somewhat shy and enjoyed watching the other classmates from a distance [3].

*Early school years*

In 1886, Mileva attended a school for girls in Novi Sad but transferred to another school in Sremska Mitrovica a year later. Four years later, in 1890, Mileva enrolled at the Royal Serbian Grammar School in Šabac. In his book *Mileva and Albert Einstein: their love and scientific collaboration*, Djordje Krstić confirms Mileva's aptitude, writing, "Mileva, among 14 pupils of her class, had excellent marks in both mathematics and physics” [3]. Interestingly, Mileva also learned German to a proficient level before 1890. From an early age, she played the tamburitza, a traditional, mandolin-like instrument, and later also played the piano [12].

*High school*

In 1891, Miloš attained special permission to enrol Marić as a private student at the exclusively male Royal Classical High School in Zagreb. John Stachel, who became the first editor of the Einstein Papers Project, has stated, “Her secondary education was quite unusual for a girl of that time and place” [13]. This young Serbian girl passed the entrance exam and entered her tenth grade in 1892. In February of 1894, Mileva was granted special permission to attend physics lectures; in September, she passed the final exams, in all subjects. Her grades in mathematics and physics were [14]. Then, Mileva fell quite ill and moved to Switzerland. After her recovery, she attended the Höhere Töchterschule (girl’s high school) in Zürich from 1894 to 1896 [14].

*University acceptance*

In 1896, Mileva successfully passed the Matura-Exam and was accepted into medicine at the University of Zürich. However, medicine was not to her liking, and in autumn of 1896, she transferred to the Swiss Federal Polytechnic in Zürich, which in 1911 became known as Eidgenössische Technische Hochschule (ETH). At that time, Switzerland was one of few countries where women were permitted to attend university. Mileva proved her academic intellect by passing the mathematics entrance examination with grade of 4.25 (on a scale of 1–6) [15]. Mileva then enrolled for the Diploma course within section VIA, which would allow her to teach physics and mathematics for secondary school students. She was the only woman in her group of six students and the fifth woman to enter the VIA section. She was 20 years old at that time. Albert Einstein had enrolled in the same program in that same year, and eventually, they became much more than classmates [3].

## 4 Early Biography—1879 to 1896: Albert Einstein

Albert Einstein was born on Friday, March 14,1879, in Ulm, in the Kingdom of Württemberg in the German Empire, which had recently joined the new German Reich. Einstein was the first child to Pauline and Hermann Einstein [12]. Pauline thought the baby was quite large and angular and believed that he might have suffered some kind of birth defect [16]. At the age of two, little Albert Einstein was referred to as *der Depperte* (or the dopey one)because he had difficulty talking. His parents even consulted a physician, they finally resigned themselves to the fact that the child was just little backwards [12]. In 1881, Einstein’s young sister Maria was born, and she was later nicknamed Maja [16].

*Early life*

The Einstein family moved to Munich in 1880. There Hermann founded an electrical manufacturing company, which was based on direct current and which he called Elektrotechnische Fabrik J. Einstein & Cie. Hermann had previously failed in a featherbed business, so this new business was a wonderful new opportunity [16].

The Einstein family was Jewish but not observant. At the age of six, Einstein attended a Catholic elementary school. Then, at the age of nine, he enrolled ina Catholic Primary School the Luitpold Gymnasium[1]. There, he received advanced primary and secondary school education from 1886–1895 until he left Germany seven years later.

The legendary story of a young Einstein being mesmerized by pocket compass, his father had given him, is entirely true; Einstein realized that there must be some unseen force involved in the working of the compass. This occurrence was one of the first signs of Einstein's curiosity and intellect. Einstein took music lessons on the violin from age six to age thirteen, often accompanied on piano by his mother [17]. Einstein's violin lessons sparked his love of Mozart. Later in life, Einstein would claim that, by the age of twelve he had mastered differential and integral calculus [12].

*Upheaval*

In 1885, Hermann's electrical business had over 200 employees and had supplied the first electrical lights for Munich's Oktoberfest. However, by 1894, at which time Einstein was fifteen years old, the Elektrotechnische Fabrik J. Einstein & Cie had failed, due to the supremacy of alternating current over direct current. The Einstein family moved to Italy seeking new business opportunities, while Einstein stayed in Munich to finish his studies. However, in late December 1894, he travelled to Pavia,Italy, to join his family. In 1895, while living in Pavia,Einstein did calculations for his uncle Jakob's new invention, concerning arc-light improvements, and astounded his uncle and the whole family [12].

*Failed acceptance to university*

In the same year, Einstein took the entrance examinations for the Swiss Federal Polytechnic in Zürich. However, he failed to attain the required standard in the general part of the examination, yet scored exceptional grades in physics and mathematics [12].

*Successful graduation*

---

[1] This school is now known as the Albert Einstein Gymnasium.

A friend of the family recommended that young Einstein attend the Argovian Cantonal school in Switzerland in the town of Aarau, near Zürich, which he did in the fall of 1895. In 1896, Einstein graduated from the Aarau school, achieving the highest marks overall in his class with the best grades in mathematics, physics, and German [16]. It was at this time, while living in Switzerland, that Einstein decided to renounce his German citizenship. Although the Nazis had not yet seized power, the anti-Semitic propaganda had blatantly emerged [18].

*First writings on light*

In the summer of 1895 Einstein sent his uncle Jakob a document that detailed the propagation of light through ether, mathematically. Jeremy Bernstein suggests, in his book *Albert Einstein: and the frontiers of physics*, "The document could have been written by any conventional 19th century physicist" [17, 19]. Einstein had written the document before he had entered university, and it clearly revealed his obsession with light.

*Successful acceptance to University*

In September 1896, Einstein passed the Matura-Exam with reasonably good grades, including a perfect mark of 6 in physics and mathematics on a scale of 1–6. Although he was only 17, this young man enrolled in the four-year mathematics and physics teaching diploma program at the Swiss Federal Polytechnic [16].

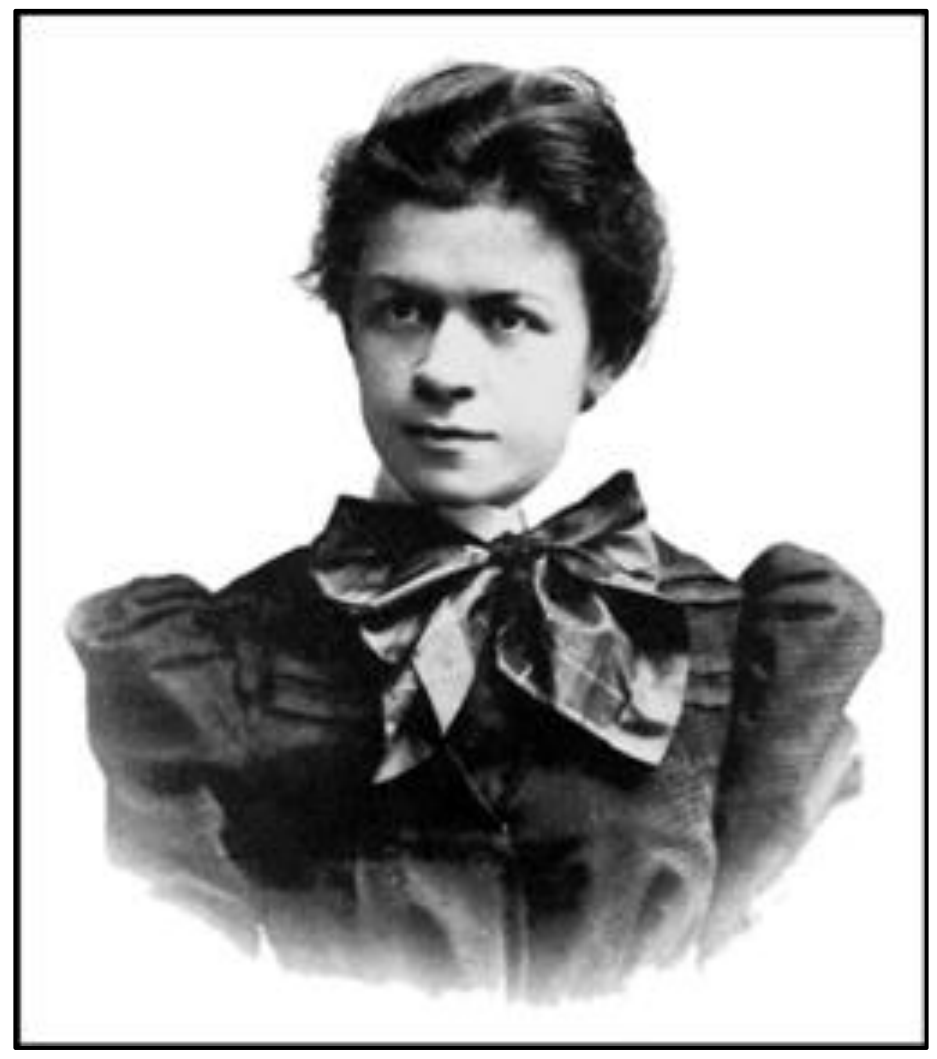

**Fig. 1** Mileva Marić. Circa 1896.
Credit: Wikipedia.
https://en.wikipedia.org/wiki/Mileva_Mari%C4%87

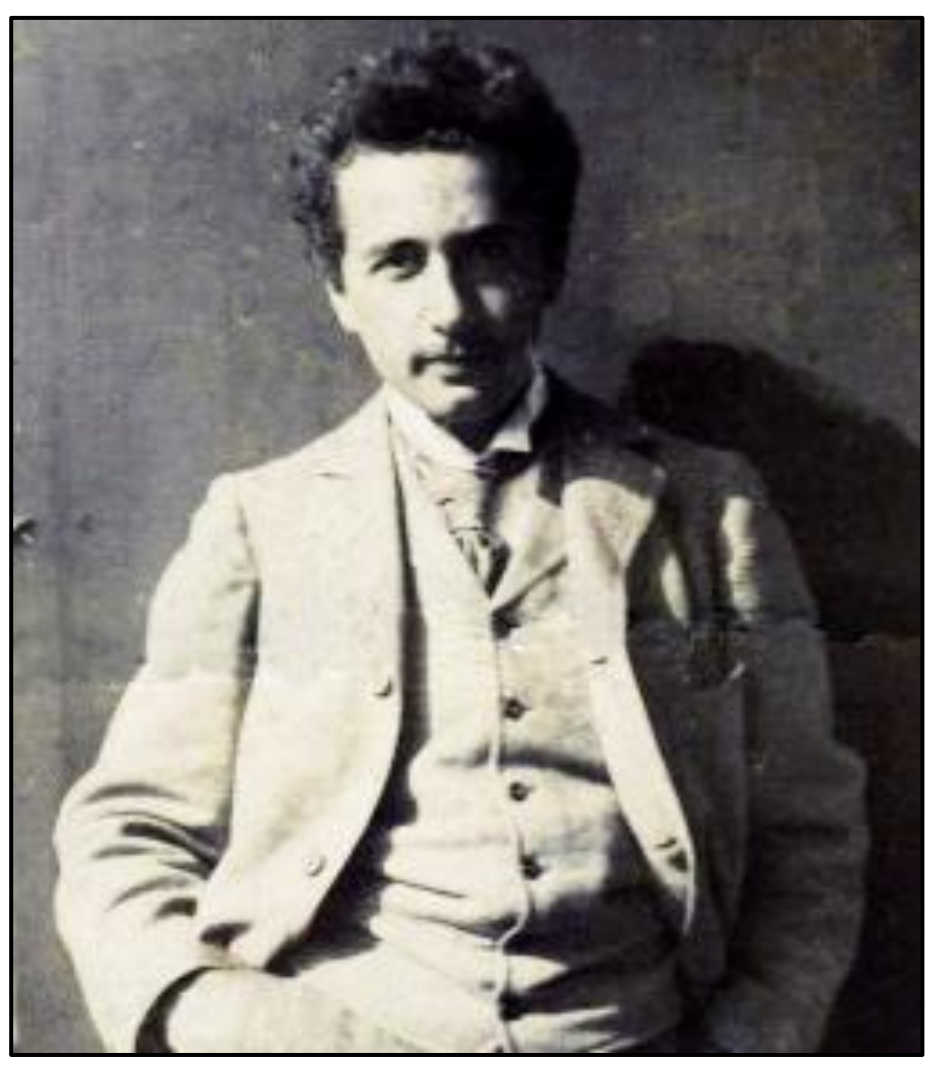

**Fig. 2** Einstein at Zurich Polytechnic.
Circa 1896. © 2015 Leo Baeck Institute.

## 5 Mileva and Einstein —1896 to 1919

In his first year at the Swiss Federal Polytechnic in Zürich, Einstein was 17 years old, and Mileva was three years his senior. It was not long before they noticed each other in lectures and soon became friends. Einstein instantly liked the reticent Serbian girl. Milan Popović describes Mileva's personality in his book *In Albert's shadow: the life and letters of Mileva Marić, Einstein'sfirst wife*: "Mileva Marić was a quick wit, a talented musician and accomplished mathematician and a promising physicist" [20]. Mileva excelled in her studies from the beginning of her time at the Swiss Federal Polytechnic. Clearly, a young woman studying theoretical physics in a European institution in the late 19$^{th}$ century would have been a thorn in the side of the conventionally stoic world of academia. However, this would not deter the young, ambitious Albert Einstein. After all, he was something of a rebel himself and believed that the Swiss Federal Polytechnic was not keeping pace with progress in physics.

*Friendship*

Over the next couple of years, Mileva and Einstein's friendship become solidified; they often read books together on new physics developments, which was not part of the course work. As a result, they spent much time together [12]. Einstein also made friends with other students at the university and seemed well liked. However, Mileva was slower to form friendships and appeared somewhat solitary. Einstein relied on a friend that would become very instrumental in his life, Marcel Grossman, who took meticulous lecture notes and lent them to Einstein on many occasions. It seems that Einstein missed lectures because he believed that the course material was not up to date, especially since the monumental electromagnetic equations of James Clerk Maxwell were not included in the course curriculum [21].

*Gasthörer*

After the October 1897, in her 3$^{rd}$ semester, Mileva attended Heidelberg University. Her professor was Philipp Lenard, who would later win the Noble Prize in 1905 [22]. Mileva was very excited about the work that Lenard was doing. She wrote to Einstein about details of experiments inOctober 20 of that year: "Oh, it was really neat at the lecture of Prof. Lenard yesterday; he is talking now about the kinetic theory of heat of gases."She continues, in the same letter, to explain the details of one of Lenard's lectures about the kinetic theory of heat of gases (Letter 36, page 34) [2]. Einstein was very interested in what Mileva was studying and replied with details of important course work at the Swiss Federal Polytechnic. Mileva was only permitted to audit the course at Heidelberg as a *Gasthörer*, but not to obtain credit, for she was female. Of particular interest to Einstein was Lenard's work on cathode rays or electrons. Hans C. Ohanian states in his book *Einstein's mistakes: the human failings of genius*, "Lenard found that the energy absorbed by the electrons from the ultraviolet light increased with the frequency of light, but it did not depend on the intensity of light" [23]. The letters, between Mileva and Einstein, during this period, mixed physics with romance. Walter Isaacson, in his lucid biography *Einstein — his life and universe*, reveals the seeds of their romantic involvement "After she moved to Heidelberg, [Mileva] shows glimmers of a romantic attraction but also highlights her self-confident nonchalance" [12].

*Returning to the Swiss Federal Polytechnic*

In April 1898, Mileva returned to Zürich and the Swiss Federal Polytechnic, and Einstein wanted to hear all the details of the lectures with Lenard. Swiss Federal Polytechnic Mileva's studies included: experimental, applied, and theoretical physics; mechanics; differential and integral calculus; descriptive and projective geometry; and astronomy. Together once again at the Swiss Federal Polytechnic, Mileva and Einstein became inseparable, and "it is easy to see why Einstein felt such an affinity for Marić. They were kindred spirits, who perceived themselves as aloof scholars and outsiders" [12]. Einstein took his intermediate exams in October 1898 and finished first in his class with an average mark of 5.7 out of a possible 6. In second place was his friend and confidant Marcel Grossman.

Einstein wrote to Mileva in August 1899 that he was enjoying collaborating with her, saying, "I find the collaboration very good" (Letter 50, page 129) [2]. Evan Harris Walker makes this particular letter out to be significant in his paper entitled "Ms. Einstein" [24] because in the letter, Einstein's focus course material but on extracurricular studies.

*Intermediate diploma examinations*

Later in 1899, Mileva took the intermediate diploma examinations; she was one year behind other students in her group because of her stay at Heidelberg University. She obtained a grade average of 5.05 of 1–6. This gave Mileva the fifth best score out of the six students in her group. In the previous year's group, Einstein had been top of the class with a grade of 5.7. Interestingly, Mileva attained the same mark in physics as Einstein had— 5.5 [3]. Cly,their capabilities in physics were evenly matched.

From 1899 onwards, Einstein, Mileva, and Mileva's good friend, Helene Kaufler often played music together [20]. Einstein and Mileva's relationship had development into one of mutual adoration, and from October 1899, they became lovers [3].

*Dissertations*

By March 9, 1900, Mileva had chosen the topic of her dissertation. In a letter to her close friend Helene, she wrote, "Professor Weber accepted my proposal for my diploma work and was very satisfied with it.

I am glad of the research, which is ahead of me. E [referring to Einstein] also accepted an interesting theme for himself" (Letter 63, page 138) [2]. Interestingly, we do not know the exact area of research to which the letter refers, but we do know that both Mileva's and Einstein'swritten works were accepted. There is also an interesting letter from Helene to her mother, Ida, of Mileva's and Einstein's progress regarding the outcome of these dissertations. Krstić paraphrases part of this letter: "Miss Marić and Mr. Einstein have now completed their written works. They planned them together, but Mr. Einstein left the most beautiful part to Miss Marić. He will probably become an assistant to his professor and will remain here. Miss Marić was also offered an assistant's post at the Polytechnic" [3]. Letters outside the couple's own correspondence, such as Helene's, reveal the public perception of their academic progress, which seems advanced for both Mileva and Einstein.

Final grading and graduation took place in 1900. Heinrich Friedrich Weber was both Einstein's and Mileva's thesis advisor, and he gave their papers the two lowest essay grades in the class, with 4.5 and 4.0, respectively. Einstein had never liked Weber and had exclaimed on more than one occasion that his lectures were 50 years out of date because they did not contain James Clerk Maxwell's equations which formulated the classical theory of electromagnetic radiation [12]. Despite Einstein's poor performance, he managed to attain a final average grade of 4.9, making him forth best in the class. However, Mileva did reasonably well in all subjects, except in the mathematics component(i.e., theory of functions), in which she scored 2.5. Because of this relatively low mark, her final grade was 4, the lowest grade in the class [2]. Einstein graduated with his diploma in July 1900, while Mileva resigned herself to retrying the examination the following year [14].

*Considered equals*

Einstein and Mileva were clearly in love and devoted much time to each other, and their mutual passion for physics strengthened their relationship. They discussed topics, read books, and became somewhat autodidact as well, studying new physics that was not part of the Swiss Federal Polytechnic curriculum [16]. Krstić states that "Mileva regularly worked, mostly in the evenings and during the nights, at the same table with Einstein — quietly, modestly and never in public view" [3]. At their time at the Swiss Federal Polytechnic, had proclaimed in letters to Mileva that he considered her "a creature who is my equal" (Letter 79, Page 152) [1].

*Wilhelm Fiedler*

Otto Wilhelm Fiedler, or simply Wilhelm Fiedler, was a Swiss-German mathematician, who had written geometry textbooks and made contributions to descriptive geometry. Fiedler taught the geometry portion of the theory of functions course. While many writers reiterate Mileva's failure in this course, very little is mentioned of the tutor who was also the marker of the paper; namely, Wilhelm Fiedler. The other students in Mileva's small group — all male — obtained at least 5.5 in the theory of functions course. It was only Mileva attained a low grade of 2.2 [2]. Wilhelm Fiedler was in that year a member of the Prussian Academy of Sciences, and some members of that body felt there was no place for women in science, let alone physics. To these men, medicine seemed like a more fitting area of specialisation for women [16]. Some researchers have suggested that Mileva merely had a poor mathematics grade and was thus incapable of obtaining the diploma. The most vehement critic of her grades is Allen Esterson [25]. Most researchers, like Esterson, fail to take into consideration the historical attitude towards women of that time in universities such as the Swiss Federal Polytechnic, with its conservative Prussian gentlemen of science. The Prussian Academy of Sciences was a stoic society, and it was not until 1964, almost a century after Mileva's time, that the first woman joined the society. Elisabeth Charlotte Welskopf was elected as the first female member of the successor organization to the Academy of Sciences, the German Democratic Republic (GDR) [26].Swiss Federal Polytechnic

*Mileva's and Einstein's ideas*

In 1901 Mileva studied rigorously,  hoping to go on to a complete a doctorate and become a physicist. Her parents had invested a great deal in her education, both emotionally and financially, so she felt that she needed to succeed, if only for them. Meanwhile, Einstein was busy looking for work. It seems no one wanted to employ this new graduate, even though all of his classmates had secured good positions [16]. Mileva and Einstein continued their study of any new physics papers. In fact, in a letter Einstein wrote to Mileva, dated March 27, 1901, He said, "How happy and proud I will be when the two of us together will have brought *our* (my emphsis added) work on relative motion to a victorious conclusion!" (Letter 94, page 161) [2, 24]. Later, in another letter, Einstein wrote to her, "The local Prof. Weber is very nice to me and shows interest in *my* (my emphasis added) investigations. I gave him *our* (my emphasis added) paper" (Letter 107, page 171) [2]. Clearly, in this letter, is talking about two different research items —*his* investigations, and his *and Mileva's* collaboration. In yet another letter, from Einstein to Mileva, , he wrote, "I am again studying Boltzmann's theory of gases. I think, however, that O.E. Meyer has enough empirical material for *our* (my emphasis added) investigation. If you once go to the library, you may check it;" and later in the same letter, "I am very curious whether *our* (my emphasis added) conservative molecular force will hold good for gases as well" (Letter 102, page 168) [2]. John Stachel and other writers have suggested that Einstein's use of pronouns was merely a romantic inflection [13]. However, there is no evidence elsewhere that the soon-to-be world famous physicist would mix up his ideas with other people's. The most natural conclusion is that he was referring to several ideas: some were his, others were theirs, and perhaps still others were solely Mileva's.

*Pregnancy*

Mileva's academic career was further disrupted in early 1901 when she became pregnant. Zürich at that time was a center for the burgeoning birth control industry, whereby advertisements offered mail order abortion drugs. Mileva could have terminated the pregnancy, but she decided to have Einstein's baby even though he was not prepared to marry [12].

*Final diploma examination*

On Friday, July 26, 1901, while three months pregnant, Mileva sat for her second attempt at the final diploma examination. However, she did not pass because of the same subject that had hurt her score the first time— the theory of functions. The examination was very stressful for the new mother-to-be. Strangely, her marks in theoretical and experimental physics were lower than they were in the previous attempt. Krstić suggests that "the stress of being secretly pregnant and not married may have affected Mileva's concentration." Krstić further suggests that Mileva's lower mark might have been related to Weber, who graded her examination: "Because of Einstein's caustic relationship with Professor Weber, Mileva was no longer was on good terms with Professor Weber" [3]. Her status as a woman might also have been a factor contributing to her low grade. As a result of her failed second attempt, Mileva discontinued her work on her diploma dissertation, which she had hoped to use as a doctoral thesis. This diploma dissertation also was under the supervision of the physics professor Heinrich Weber.

*Swiss citizenship*

While Mileva failed her exam, Einstein was awarded Swiss citizenship. Since renouncing his German citizenship some years earlier, he had been stateless; this acceptance into Switzerland as a resident afforded him a sense of belonging.

*Physics*

Many critics have said that Mileva never wrote about physics to Einstein. In a letter dated early November 1901, Mileva wrote to Einstein, "What nice books you sent me ... I've also read the one by Forel; when I finish it, I'll write to you about it. Have you read the one by Planck? It seems to be interesting" (Letter 123, page 182) [2]. The real issue here is that many of Mileva's letters to Albert Einstein cannot be found. Some authors suggest that she wrote many letters to him about physics, but we are not privy to them [26]. The reason that the letters are missing remains a mystery.

*First scientific paper*

On December 13, 1900, submitted his first real scientific paper, “Folgerungen as den Kapillaritätserscheinungen” (“Conclusions drawn from the phenomena of capillarity”), to the prestigious *Annalen der Physik*[2]. It was published in March of 1901 [27]. It was a time of great excitement for Einstein and also for Mileva. It is clear that saw this work as the product of their collaboration. On April 4, 1901, Einsteinwrote to Mileva, “He [Michele Besso] is very interested in *our* (my emphasis added) investigations ... The day before yesterday, he [Michele Besso] went on *my* (my emphasis added) behalf, to see his uncle, Prof. Jung, one of the most influential professors of Italy and also gave him *our* (my emphasis added) paper” (Letter 96, page 162) [2]. The letter was referring to the paper “Conclusions drawn from the phenomena of capillarity,” which Einstein wrote was a product of their collaboration. In fact in the documentary *Einstein’s wife*, Robert Schulmann, who was an editor of *The collected works of Albert Einstein* and who worked alongside John Stachel, says of Mileva’s involvement, “It is very conceivable that Mileva had input on the paper, on Capillarity (referring to Einstein’s paper, “Conclusions drawn from the phenomena of capillarity”). That, of course, has nothing to do with special relativity. However, it is fair enough to say that Mileva could conceivably have contributed to the first paper of his” [28]. Other writers vehemently disagree, Esterson being the most vocal [25].

*Relative motion work*

In another letter from Einstein to Mileva December 19, 1901, he wrote, “Today I spent the whole afternoon with Kleiner [Alfred Kleiner] in Zürich and explained *my* (my emphasis added) ideas on electrodynamics of moving bodies to him and otherwise talked about all kinds of physical problems” (Letter 130, page 189) [2]. Einstein switched doctoral advisor, from Weber to Alfred Kleiner. We see here in this letter, that is taking a departure from the collaboration. Although he refers to the work prior to this time as *theirs*, even the relative motion work, he is clearly adding more to the idea and is now starting to see it as *his* idea.There seems to be a dichotomy here between what is collaborative and what is individually novel for . In the same letter, he adds, “I will certainly write the paper in the coming weeks” [2]. Many writers, especially Stachel, have iterated the point that this is evidence that Einstein was only referring to “our” ideas in the romantic sense [13].

However, I propose an alternative hypothesis here, namely, that had been considering two sets of ideas, one he developed with Mileva and a second he developed himself. The letter of December 1901 indicates that he preferred *his* ideas over *their* ideas. This seems more in line with the theoretical process when developing a concept to explain physical phenomena and is more natural explanation. As Stachel was so influential regarding the interpretation of the letters, other authors merely reiterated the *sounding board* idea, so much so that even a recent article of 2012 perpetuates Stachel's sounding board idea, which is evidently a misinterpretation [29].

---

[2] *Annalen der Physik* (*Annals of Physics*) is one of the oldest scientific journals on physics, founded in 1799.

*Unwanted baby*

Mileva's pregnancy had subdued her excitement for becoming a physicist. By 19th century standards, her child was illegitimate and unwanted. This most certainly caused heranxiety, Einsteinespecially since his family had shown disdain for her in every possible way.

In late January of 1902, Mileva gave birth to first child, Lieserl. Einstein never visited Serbia to meet the baby. Moreover, he certainly didn't want his family finding out about the illegitimate child. Lieserl is believed to have had scarlet fever and lived at Mileva's parent's home in Novi Sad. In his biography about Einstein, Walter Isaacson writes "when Lieserl turned 19 month of age, she was given up for adoption" [12]. However, in Michele Zackheim's book *Einstein's daughter: The search for Lieserl*, the story is more complex and compelling. Zackheim, who researched Lieserl's short, poignant life, suggests that she died of scarlet fever, since no record of her existence appears after the child's birth record [30]. It would be 30 years after the child's birth before the truth of Einstein's first child would surface publicly.

*What Einstein considered collaborative and his own*

On February 8, 1902, while Mileva was still at Novi Sad, Einsteinwrote to her, "I am now expounding to Habicht [his friend Conrad Habicht] the paper I have handed into Kleiner [his new Ph.D. supervisor]. He is quite enthusiastic about *my* (my emphasis added) good ideas and nags me to send to Boltzmann [Ludwig Boltzmann] the part of the paper that refers to his [Boltzmann's] book" (Letter 136, page 192) [2]. made a clear distinction here, between what he considered to be *his* workand what he considered to be *their* collaborative work.

*Second scientific paper*

In April 1902, Einstein published his second paper in *Annalen der Physik,* "Thermodynamische Theorie der Potentialdifferenz zwischen Metallen und vollständig dissoziierten Lösungen ihrer Salze, und eine elektrische Methode zur Erforschung der Molekularkräfte" ("On the thermodynamic theory of the difference in potentials between metals and fully dissociated solutions of their salts and on an electrical method for investigating molecular forces") [31]. A few years later, in 1907, Einstein would say that this paper and the previous one were worthless and irrelevant.

*Akademie Olympia*

On Easter day of 1902, a group of Einstein's friends, mostly from Bern in Switzerland formed an intellectual society of sorts. It was half discussion forum and half dinner party. The name of the gathering was Akademie Olympia (the Olympia Academy). They usually met at  apartment and discussed philosophy, mathematics, and physics. Mileva attended most meetings, but as Maurice Solovine, another member of the group, later recalled, "Mileva, intelligent and reserved, listened intently but never intervened in our discussions" [12]. Aside from listening, Mileva also took notes of everything that was said.

The founding members of the Akademie Olympia were Einstein, Conrad Habicht, and Maurice Solovine. Solovine was a Romanian philosophy student who answered a newspaper advertisement concerning physics tutorials that had published. Soon after, a mathematician by the name of Conrad Habicht joined the group, which by that time become Akademie Olympia. Solovine and Habicht became lifelong friends of  [12]. There were other people who participated in some meetings. Paul Habicht, the brother of Conrad Habicht, came to a few. Michele Besso, a mechanical engineer, Marcel Grossmann, a mathematician and Einstein's friend at Swiss Federal Polytechnic, and Lucien Chavan, an electrical engineer, also joined in periodically [12].

Now, not only Mileva studied physics and mathematics at two prestigious European universities, but she had also attended regular meetings of a think-tank that would be instrumental in Einstein's later scientific publications. In addition, according to Krstić, during her evenings alone with Einstein, Mileva collaborated with him on scientific problems [3]. However, the notes that Mileva took at such meetings and the evenings with Einstein would definitely reveal much of her involvement, but all her work has never been found.

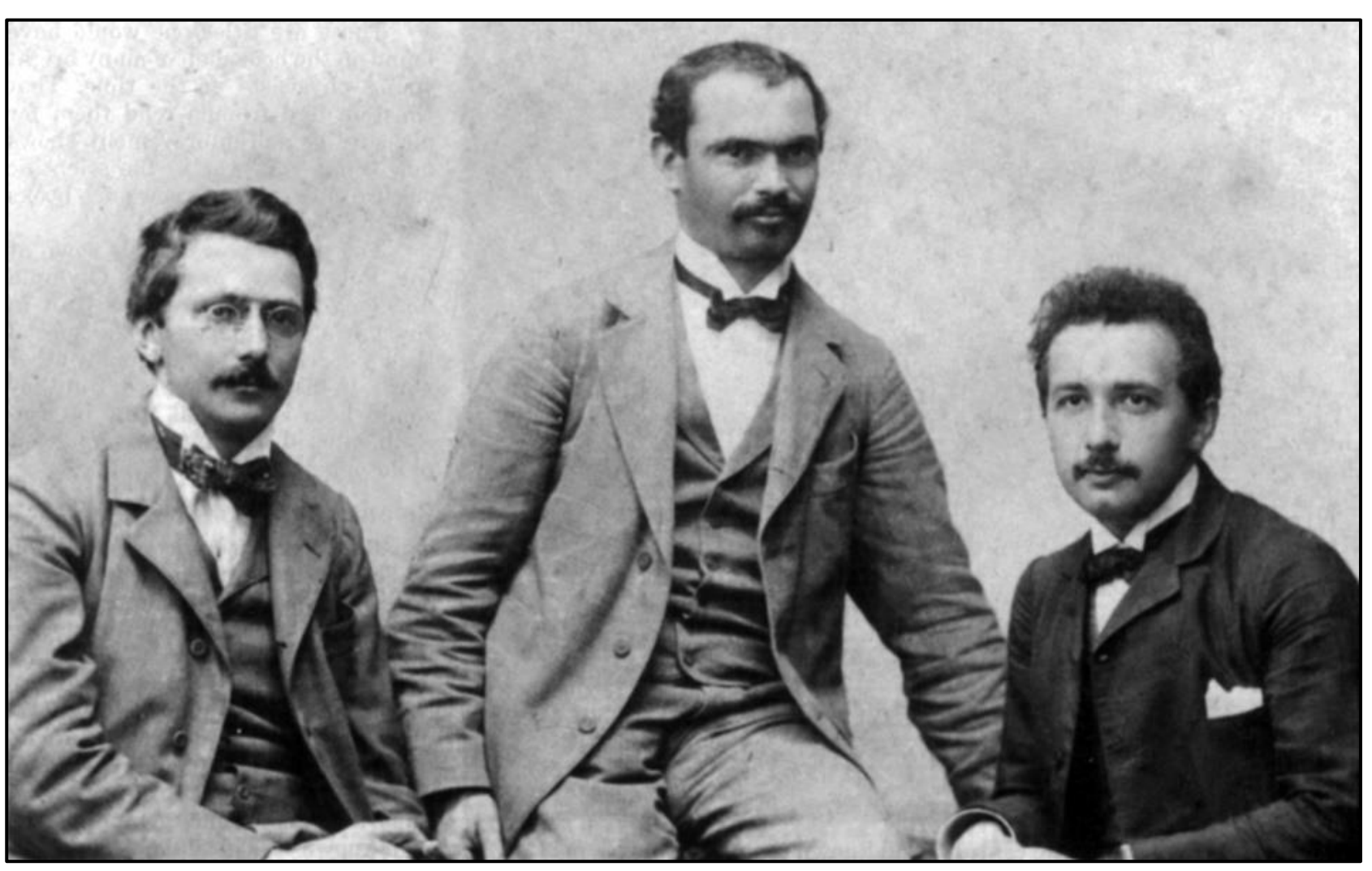

**Fig. 3** Akademie Olympia. Left to right: Conrad Habicht, Maurice Solovine, and Albert Einstein. Circa 1903. Credit: Wikipedia. https://de.wikipedia.org/wiki/Akademie_Olympia

*Third scientific paper*

In June 1902, published his third paper in *Annalen der Physik*, "Kinetische Theorie des Wärmegleichgewichtes und des zweiten Hauptsatzes der Thermodynamik" ("Kinetic theory of thermal equilibrium and of the second law of thermodynamics") [32]. As with the previous paper, this paper pertains to some of the topics of discussion in the correspondence between Mileva and Einstein. However, as in the previous papers, only Einstein's name appeared on the document, and that would continue to be the case with all of Einstein's scientific papers of that period.

*Patent clerk position*

Einstein had been living hand-to-mouth since leaving Swiss Federal Polytechnic in 1900. In December 1901,he heard of a patent clerk job that was available in Bern [12]. On June 16, 1902, the job was offered to Einstein. Albert Einstein was now employed as the lowest rank patent clerk, provisionally as a Technical Expert Class 3. He would remain a patent clerk till 1909.

Mileva was happy that Einstein now had work; it gave their relationship the security it needed, and it allowed them to continue their studies and to hold meetings of the Akademie Olympia. Even though Mileva was no longer studying at a university, her interest in physics continued, as did her love for . They became inseparable. There is no doubt that Akademie Olympia played a significant role in 'intellectual development, leading up to his 1905 *annus mirabilis*. Up until the time of their marriage, Mileva was immersed in all of Einstein's activities and underwent a similar intellectual development. They were happy days for the young couple, filled with fond memories [16].

*Marriage of Mileva and Einstein*

On January 6, 1903, Einstein and Mileva married at the Bern registrar's office in a tiny civil ceremony. Their Akademie Olympia colleagues, Maurice Solovine and Conrad Habicht, served as witnesses to the wedding. There were no family members from either side, only their regular group of intellectual confidants, who later that evening dined with them at a restaurant as a small reception.Mileva did not have a maid of honor or any bridesmaids, but she and Einstein were both happy simply with the presence of their two intellectual friends, who were, in fact, the only attendants of their wedding. There was no honeymoon; the couple quietly returned to their apartment together that night [12].

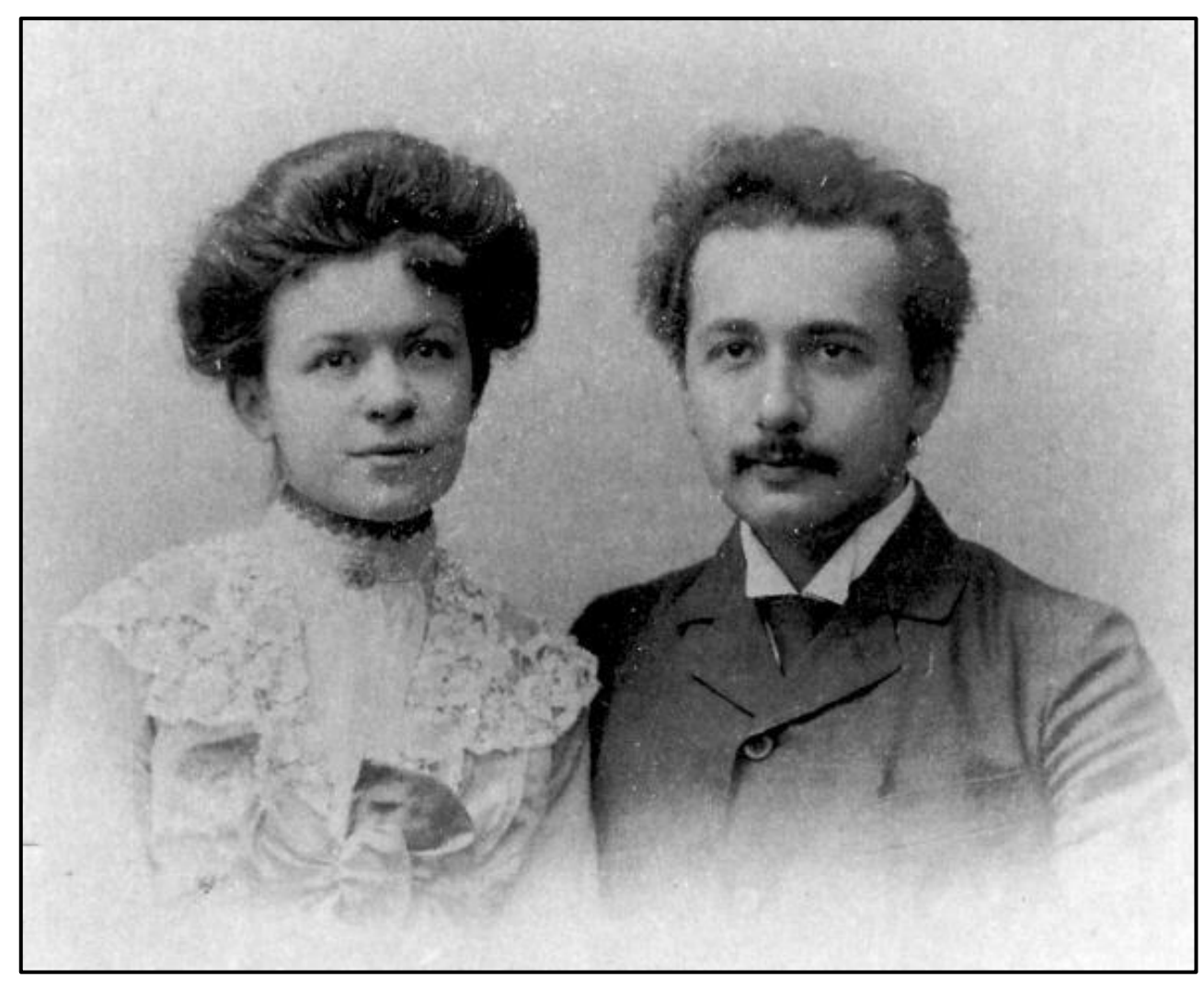

**Fig. 4** Mileva and Albert Einstein on their wedding day.

Credit: Tesla Memorial Society.

*Life in Bern*

In letters to friends, both Mileva and Einstein wrote of their happiness and love. Life in Bern become very comfortable for them. Mileva wrote to a friend, "I am even closer to my sweetheart if it is at all possible than I was in our Zürich days" [12].

Sadly, however, Mileva eventually became unhappy. She felt as if she had become a mere onlookerduring scientific discussions [20]. Perhaps this unhappiness was related more to her personality her academic ability, for she was known to be withdrawn at times and also suffered from depression.

*Fourth scientific paper*

In Janurary 1903, Einstein published his fourth paper in *Annalen der Physik*, "Eine Theorie der Grundlagen der Thermodynamik" ("A theory of the foundations of thermodynamics") [33]. This fourth paper, like the three before it, was statistical in its mathematical form and bore a heavy dependence on the sigma notation.

*Fifth scientific paper*

In March 1904, Einstein published his fifth paper in *Annalen der Physik*, "Allgemeine molekulare Theorie der Wärme" ("On the general molecular theory of heat") [34]. Up until now, Einstein's papers had gone unnoticed, but that would not always be the case. However, it would take years before the real value of these papers would be understood.

*First son is born*

Then, in May 1904, the Einstein's first son, Hans Albert Einstein,was born in Bern. It was a happy event, and it gave Einstein and Mileva's marriage a much-needed boost. Mileva had a new purpose in taking care of Hans Albert, and it must have meant a great deal to her after the situation with her firstborn, Lieserl. The couple welcomed Hans Albert's arrival, which rekindled a spark in their lives. From a modern standpoint, Mileva might have been classified as suffering from chronic depression, a condition that would appear again, sooner or later [16].

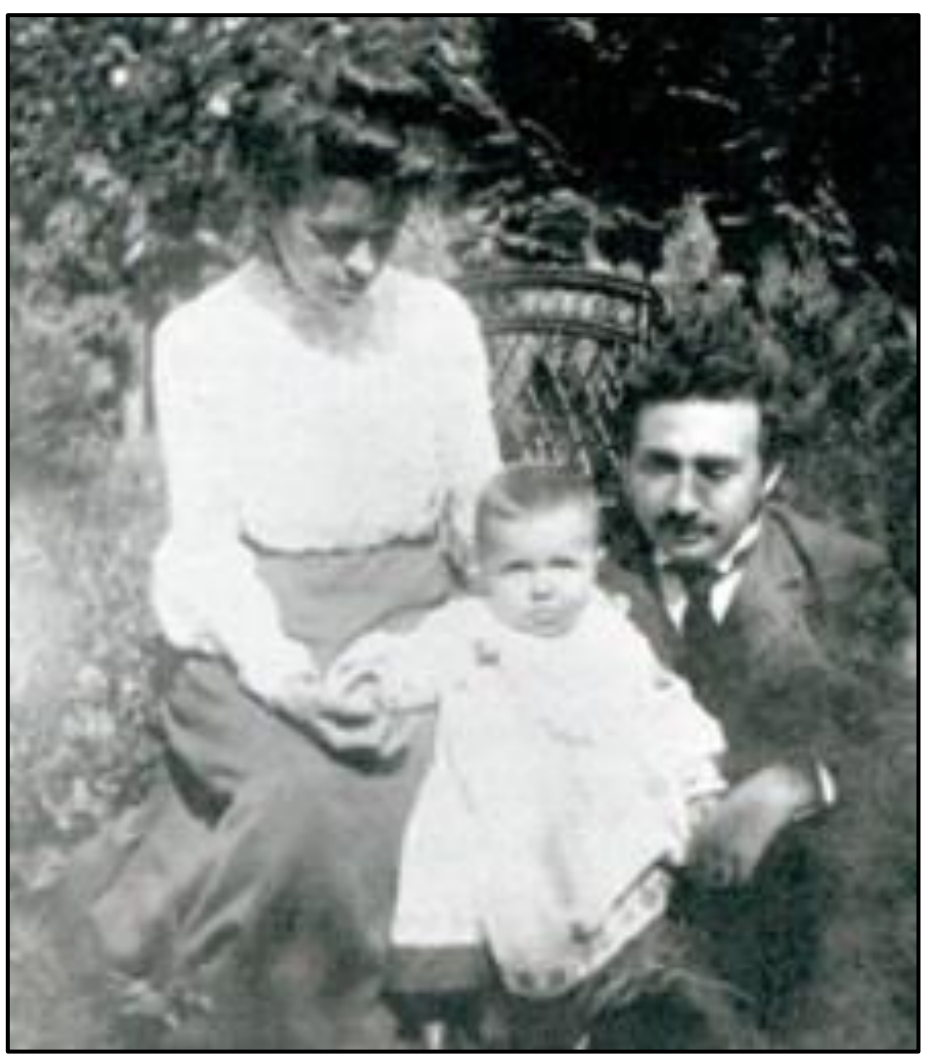

**Fig. 5** Mileva and Einstein with Hans Albert. Circa 1904.

Credit: Tesla Memorial Society.

*Doctoral thesis*

On April 30, 1905, Einstein submitted his completed thesis to Alfred Kleiner, professor of Experimental Physics. The doctoral dissertation was entitled, *A New Determination of Molecular Dimensions*. Shortly afterwards, Einstein was awarded a Ph.D. by the University of Zürich [35]. Interestingly, this is the same work that Einstein and Mileva had corresponded about in earlier letters. Moreover, some of the work related to topics that Mileva had also researched. There is a strong possibility that her thesis work was on a similar topic.

*Annus Mirabilis*

1905 is referred to as Einstein's *annus mirabilis* because he published four monumental papers in physics. These were fundamental pieces of work that he had apparently been working on for some time. Although they were released only a month or two apart, sequentially, they were clearly the product of many months or even years of work.

The sixth scientific paper and the first one in 1905, submitted in March, was "Über einen die Erzeugung und Verwandlung des Lichtes betreffenden heuristischen Gesichtspunkt" ("On a heuristic point of view concerning the production and transformation of light") [36]. This paper was about photoelectricity and described how light consists of photons or particles of a given quantum of energy. This landmark paper would later win Einstein the Nobel Prize in 1922 [37]. The work that Mileva had done at Heidelberg University with Professor Lenard was instrumental in the forming Einstein's paper on photoelectricity because it allowed Einstein to calculate the fundamental value of a photon [3]. Many critics have insisted that Mileva took no part in any of Einstein's work. However, there is evidence that Mileva had prepared letters to Max Planck about this work on photoelectricity, so apparently she was involved [3].

In May Einstein, submited his seventh paper to *Annalen der Physik*, "Über die von der molekularkinetischen Theorie der Wärme geforderte Bewegung von in ruhenden Flüssigkeiten suspendierten Teilchen" ("On the movement of small particles suspended in stationary liquids required by the molecular-kinetic theory of heat") [38]. This was a seminal work concerning the Brownian motion of molecules and was, again, another subject about which Einstein and Mileva had corresponded [1].

Then, merely one month later, in June, Einstein submited his eighth scientific paper. It was another landmark work, published in *Annalen der Physik*, entitled "Zur Elektrodynamik bewegter Körper" ("On the electrodynamics of moving bodies") [39]. The paper was, essentially, Einstein's publication on the special theory of relativity. Einstein put an acknowledgement at the end of the paper that reads, "In conclusion, let me note that my friend and colleague M. Besso steadfastly stood by me in my work on the problem here discussed and that I am indebted to him for many a valuable suggestion" (Document 23, page 171) [40]. There is no mention of Mileva in the acknowledgement.

In September, Einstein published his ninth paper in *Annalen der Physik*, entitled, "Ist die Trägheit eines Körpers von seinem Energieinhalt abhängig?" ("Does the inertia of a body depend upon its energy content?") [41] In this paper, he showed that mass is equivalent to energy, and vice versa, by producing a reworking of an earlier Henri Poincaré equation [43], which has become the most famous equation in history; namely: $E = mc^2$.

*The waiting game*

The revolution in physics that Einstein had set in motion would not happen right away. In fact, it would take years for someone to read and understand papers, especially the ones published in 1905. n continued tirelessly to work on his ideas while at the patent office. Incidentally, he had been promoted in late 1904 to a 2nd class clerk position. He found the job ideal, for it gave him time to think about his theories.

According to Hans C. Ohanian in his book, *Einstein's mistakes: the human failings of genius*, by the end of 1905, "Mileva had lost all interest in physics; she had become Einstein's wife and cook and the mother of his children, but not his collaborator" [23]. However, in 1905, Einstein and Mileva only had one child — Hans Albert — and if Ohanian is wrong about the number of their children at that time, he is most likely wrong about other things regarding the couple. In letters from Mileva to Helene, during 1905, there is evidence that physics was still a subject that made her passionate. However, she was saddened because she felt excluded from academic life [3]. Stachel et al. point out that some letters from the period show Mileva excited and happy about her life with Einstein, which suggests that there is no reason to suspect that Mileva has lost interest in physics [40].

Some letters from Mileva to her good friend Helene during early 1907 and mid-1908 have not survived. Since this was a crucial period, it unclear to what extent Mileva collabourated with Einstein. Additionally, these letters would have provided some insight into Mileva's state of mind and would have given more details about the events that were transpiring. However, most of those letters have been lost [20].

*Max Planck*

After several years of being a patent clerk, Einstein's work was finally starting to get noticed. All the while, he continued to publish more and more papers. Then Professor Max Planck began to build upon Einstein's work. As Isaacson writes, "By the beginning of 1908, even as such academic stars as Max Planck and Wilhelm Wien were writing to ask for his insights. Einstein had tempered his aspirations to be a university professor. Instead, he had begun, believe it or not, to seek work as a high school teacher" [12]. By now Einstein was also discussing his ideas with his friend Marcel Grossmann, but he still had some physics discussions with Mileva [44]. Towards the end of 1908, Einstein was considered to be a leading physicist and was appointed lecturer at the University of Bern.

*Academia*

In 1909, Einstein gave a lecture on electrodynamics and the relativity principle at the University of Zürich, after which Alfred Kleiner recommended Einstein to the faculty in a newly created professorship in theoretical physics [12]. It was during this period that Mileva prepared notes for Einstein's physics lectures since Einstein was very busy, and she very much enjoyed being involved. She even wrote letters to other physicists on Einstein's behalf [3].

Then, in July of 1910, their second son was born in Zürich. His name was Eduard, and they nickname him Tete, for he was petite in stature. It was a very busy time for the Einsteins because Einstein's career was exploding with success upon success. In the evenings, Mileva and Einstein worked on his papers and lectures together, sitting at the same table [3]. Yet some writers still insist she was merely a housewife. If this was so, she was an exceptionally skilled housewife indeed.

*Academic notoriety*

Einstein had been in Zürich less than six months when he received an invitation to consider a more prestigious job: a full professorship at the University of Prague. In March 1910, just before relocating to Prague, Hendrik Antoon Lorentz[3] invited the Einsteins to visit him in the little Dutch town of Leiden. Mileva accompanied Einstein, and they stayed with Lorentz and his wife. Mileva was overjoyed by the invitation.

In April of 1911, Einstein became a full professor at Charles–Ferdinand University in Prague. He was also given Austrian citizenship in the Austro–Hungarian empire. In October, he attended the famous Solvay Conference, together with all the greats in physics of the day [12]. This meant that Einstein had been recognised as a leading theorist, and his life would never be the same.

*Growing apart*

Einstein was a celebrity and travelled around Europe giving lectures and enjoying his newfound fame. All the while his wife stayed in Prague, a city she hated. Mileva was depressed for not being part of scientific academia, as she had dreamt, and for having to spend most days alone with the children. Einstein started to resent his wife's demands for attention, for he was far too busy to occupy himself with family life [16].

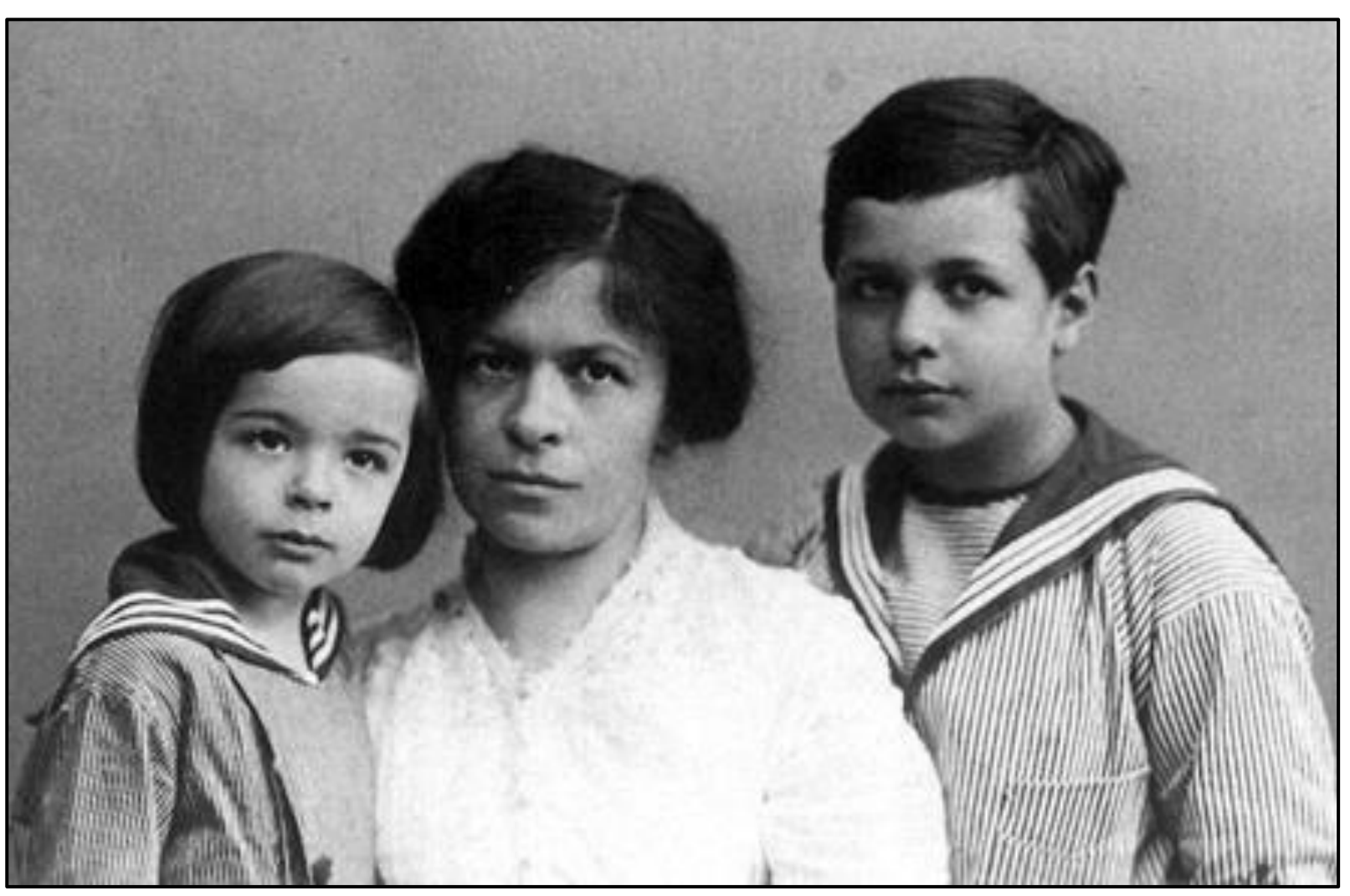

**Fig. 6** Tete, Mileva, and Hans Albert.

Circa 1912. Credit: Tesla Memorial Society.

[3] The Lorentz transformation equations were used in Einstein's Special Theory of Relativity paper in 1905.

The marriage had become unstable, for Einstein wanted the freedom to take every opportunity presented to him. On a trip to Berlin in 1912, during the Easter holidays, he traveled alone. It was at that time that Einstein became reacquainted with a cousin whom he had known as a child. Her name was Elsa Löwenthal, and she, like Mileva, was three years his senior [12]. Some authors suggest that Einstein and Elsa began an affair at that time [45].

*Full circle*

In July 1912, moved back to Zürich to take up the post of Professor of Physics for the Swiss Federal Polytechnic, which had been renamed the Zürich Eidgenössische Technische Hochschule (ETH) in 1911. He had come full circle now. Mileva was thrilled and believed that returning to Zürich could help save their marriage, as well as her sanity. The children seemed happy to be out of Prague. Einstein sent a postcard to a friend around the time of his return to Zürich that read, "Great joy about it among us old folks and the two bear cubs" [12].

However, Mileva's depression returned, and her health started to decline. She developed rheumatism, which made it very difficult for her to go outside, especially in the icy winter. At times, she was almost paralyzed by both mental and physical pain [16].

In 1914, Mileva and Einstein separated. Einstein relocated to Berlin and took the post of Professor of Physics at the University of Berlin, while Mileva and the children remained in Zürich. The couple became estranged, and Mileva had no involvement in Einstein's life from then onward. The only discussion they had was about the children and their eventual divorce. Einstein remained in his post in Berlin until 1932. In addition to his professorship, in 1917, he was appointed Director at the Kaiser Wilhelm Institute in Berlin [12]. Any collaboration between Mileva and Einstein, either emotionally and intellectually, was most certainly over.

*Divorce*

On January 31, 1918, Einstein wrote to Mileva in a letter and offered her his inevitable Nobel prize money in exchange for a divorce settlement, in which he would admit adultery. Einstein writes, "The capital would be deposited in Switzerland and placed in safe-keeping for the children" (Doc. 449, page 456) [46]. But Einstein was not true to his word; as Hans C. Ohanian points out, "After he [Einstein] had received the prize money, he gave half to Mileva [in two installments, separated by several years], but he never gave her the other half" [23]. This half compensation raises questions. First, why did Einstein use his Nobel prize money and not offer regular family support or alimony, which would be the usual thing to do? Second, why did he send only half the money? It would seem that Einstein used half the money as a sort of bribe to get Mileva to divorce him so that he could move on to a more exciting life without Mileva and the children tying him down. Or was it half the money, for half the work?

Einstein and Mileva divorced on February 14, 1919, after having lived apart for five years. An ironic date indeed. Einstein then married his cousin Elsa Löwenthal, with whom he had reignited a romance some years before. Einstein would have very little to do with Mileva for the rest of his life.

**6 Antagonists and Protagonists of the Collaboration Hypothesis**

Apparently, most of Einstein's papers between 1902 and 1905 were the product of his work since 1899. This period of 1899 to 1905 is a significant time, and if any collaboration occurred between Mileva and Einstein, it most certainly would have been during that period.

*Esterson*

There are many critics of the idea that the two worked together, in spite of the references to "our work" and "our investigation" in Einstein's letters. The most prolific antagonist to the idea that the Einsteins collaborated is Allen Esterson. In fact, Esterson's leading publications have been to oppose the idea that Mileva had in any way contributed to Einstein's work. Most of his articles that refute Mileva's contribution to Einstein's work can be found on his website [25]. Although Esterson has a Bachelor of Science, his criticisms seem to be that of a non-specialist in that he questions historical facts as opposed to facts based on physics. Esterson has not published any substantive work, and is primarily a critic who directly criticises the work of others. His main criticism of Mileva is that she failed, or so its claimed, the theory of functions course, despite the fact that all her other studies were elegant. Even if she did fail that single topic, it would not be an indication of her worthlessness. After all, Einstein failed to enter the Swiss Federal Polytechnic on his first try, yet this did not prevent him from being one of the greatest physicists in history.

*Stachel*

John Stachel, Ph.D., is an American physicist and philosopher of science. He became the first editor of the Einstein Papers Project. He was editor of the first two volumes of, *The collected papers of Albert Einstein*. He has been very critical of the Mileva collaboration hypothesis. His criticism primarily targets physics issues, so his remarks appear to be mainly that of a specialist. However, he also delves into semantics, arguing that Einstein's use of pronouns (e.g., "we," "our") should not be taken literally.

There are many other critics of Mileva's involvement, but Esterson and Stachel provide a broad cross section of academic criticism regarding the issues involved.

*The Maschinchen*

There is some evidence, that Mileva was involved with collaboration work with other scientists. In her book called *Love, Power, and Knowledge: Towards a Feminist Transformation of the Sciences*, Hilary Rose claims that "Mileva, through the collaboration with a mutual friend, Paul Habicht, constructed an innovatory device for measuring electrical currents" [47]. The device was supposedly an electrostatic generator, and it was known as the *Maschinchen* (the little machine). Esterson has criticised Rose's account, saying that it is based on an article in Troemel-Ploetz's translation of a passage from Trbuhović-Gjurić's book, *Im Schatten Albert Einsteins. Das tragische Leben der Mileva Einstein-Marić* (*In the shadow of Albert Einstein: the tragic life of Mileva Einstein-Marić*), which is supposedly devoid of any source reference [25]. However, as illustrated in this paper, Mileva helped Einstein with his work, so it is possible that Mileva also worked in collaboration with Habicht.

*Michelson-Morley experiment*

Evan Harris Walker, Ph.D., (1935–2006) was a physicist and one of the most prolific writers who suggested that the basic ideas of relativity came from Mileva. Walker published a little over a hundred papers in reputable scientific journals and held over a dozen patents. He was viewed as a radical. Stachel was one of his biggest critics, followed more recently by Esterson. In a letter to *Physics world* in February 1989, Walker wrote a lengthy piece entitled "Did Einstein espouse his spouse's ideas?" He goes on to criticise an early biography of Einstein written by Ronald W. Clark, which was considered to be an authoritative biography at the time of its publication [48, 45]. In the biography, Clark suggests that Mileva had "little learning." However, Walker knew that Mileva had virtually the same education as Einstein, so after reading Clark's biography of Einstein, he began his search for the truth.

In another edition of *Physics world*, Stachel replied to Walker's article with a scathing attack, suggesting that Walker's claims were worthy of a Hollywood screenplay and nothing more. However, in a counter response, Walker wrote an article entitled "Mileva Marić's relativistic role," in which he makes a decent argument for Mileva's involvement in Einstein's work. [49] Einstein emphatically claimed for most of his life that he was unaware of the Michelson–Morley experiment[4] when he was writing the Special Relativity paper. Stachel suggested that several letters between Einstein and Mileva showed he *did* know about the experiment.

---

[4] The experiment was an interferometer that shown essentially that light moving in the proposed luminiferous ether could not be detected. It led Hendrik Lorentz to develop the transformation equations.

However, Walker suggests that, on careful reading of the letters, it is clear Einstein had a general knowledge of the experiment but was unaware of the details, even though he used the Lorentz transformation equations. One possible conclusion, then, is that that there was part of Einstein's work on Special Relativity that only Mileva had access to. This point has never been successfully disputed, and it is one of the main arguments that Walker used to make his case for collaboration [49].

*Mileva corrected Einstein's math*

Dr. Desanka Trbuhović-Gjurić (1897–1982), a Serbian author, wrote a biographical work in 1969 in her native language, entitled *Im Schatten Albert Einsteins. Das tragische Leben der Mileva Einstein-Marić* (*In the shadow of Albert Einstein: the tragic life of Mileva Einstein-Marić*) [42]. In the work, Trbuhović-Gjurić says that when Einstein addressed a group of Croatian intellectuals, he stated, "I need my wife as she solves all the mathematical problems for me" [42]. The biography was republished in German in 1983 and received a wider audience. The book has received considerable criticism, but it is easy criticize such a work decades after it is published. The early date of the first edition adds some weight to its authenticity since it would have been closer to the source time.

Senta Troemel-Ploetz, a German linguist, says that "Einstein's ideas may have been his own, but Mileva did the mathematics in many cases" [50]. Troemel-Ploetz goes on to argue that the letters Mileva wrote to Einstein are largely not about physics but are emotional letters. This does not mean she was not involved in the physics and mathematics; it just means Mileva preferred to write in an emotional context. Troemel-Ploetz goes on to cite third-hand reports that Mileva was the workhorse in the relationship. However, Highfield and Carter describe this claim as Serbian home-town folklore and say that Mileva's title of "the Serbian Marie Curie" was unfitting [14].

In a book by Charles S. Chiu and Edith Borchardt, entitled *Women in the shadows*, the authors reiterates an important comment by Stachel that nothing novel was contributed by Mileva [51]. Chiu and Borchardt state, "The mathematics involved does not go beyond elementary calculus, and it seems unlikely that Maric contributed unique mathematical expertise … one may speculate that she might have suggested the method of proving individual results and/or checked calculations" [51]. What is interesting here is that the mathematics in the photoelectric paper and Special Relativity Paper do not go beyond elementary calculus and are indeed quite simple for someone trained in physics. However, it is not the mathematical simplicity that makes those papers monumental; it is the applied physics that does. Therefore, Stachel's original argument, that Mileva was merely a sounding board, only reinforces the idea that Mileva did do some of the mathematics [13].

*Einstein-Marity*

In 1955, Soviet physicist Abraham Joffe (1880 – 1960) published a controversial article entitled "In remembrance of Einstein." The piece was an obituary for Einstein in a leading Russian journal of physics [15]. Joffe says in the article that he saw the original three submission papers of 1905 and said they were signed "Einstein−Marity." Marity is another form of Marić. However, the name was removed from the final publications.

In response to Joffe's claim, many authors have stated that the matter was a legal one and that the quotation that the papers were signed "Einstein−Marity" was taken out of context. Alberto A. Martínez writes in his book, *Science secrets: The truth about Darwin's finches, Einstein's wife, and other myths* that, "Joffe did not claim that Marić co-authored or collaborated in any of Einstein's papers" [52]. It is believed by most, including Esterson, Stachel, and others, that the original submission papers have the name of the wife and husband as part of Swiss custom. However, the PBS Ombudsman points out in a rebuttal to Esterson and Stachel that Einstein never signed his name as Einstein−Marity in any of his previously published papers. Furthermore, "Swiss law permits the male, the female, or both, to use a double last name, but this must be declared before the marriage. Moreover, it was Mileva, not Einstein, who opted for the last name usage, Einstein−Marity" [28]. All these arguments indicate that the initial submission of three of the significant papers of 1905 were submitted by Mileva, or at the very least both the Einsteins.

*Letters and the original early papers*

It is unfortunate that many of Mileva's letters to Einstein are missing. It is also a tragedy that Einstein's initial drafts of all papers up to the end of 1905 are also missing. It seems that many of these early documents, held by Einstein, have indeed been lost or destroyed. Einstein would have known the value of those letters and first draft documents, for he knew the value of the work. So why didn't he take efforts to keep them safe? Some have speculated that he destroyed them on purpose to cover up the scientific collaboration. However, I will not cite conspiracy theories, but merely mention it as a possibility. It is equally likely he lost them or threw them away.

## 7 Notation Used in Einstein's papers

If we examine the notation employed in Einstein's early papers, particularly up to and including 1902−1905, we see a standard form of notation. This form is characterised by heavy dependence on the use of sigma notation, preceding the associated integral notation. The early equations are based on statistical analysis of physical behaviour. This type of descriptive mathematical treatment, as opposed to an implied treatment, is prevalent in most early papers prior to early 1905, but is not prominent in the Special Relativity Paper of 1905.

The mathematical treatment of special relativity does not require integration, so there is no need for sigma notation. However, the mathematical procedures of all papers prior to the 1905 that preceded "Über einen die Erzeugung und Verwandlung des Lichtes betreffenden heuristischen Gesichtspunkt" ("On a heuristic point of view concerning the production and transformation of light") have descriptive step-by-step treatments. However, following this paper, there is a trend towards an implied mathematical progression. This means that individual steps are assumed and thus implied.

In contrast, later papers concerning the special theory of relativity use sigma notation to a lesser extent and mostly in the implied form.

## 8 What Are Collaboration and Co-authorship?

In the days of Einstein, scientific authorship was much simpler and was specifically traceable to particular individuals. As physics has grown more complex, joint or multiple authored journal articles have increased. What constitutes authorship has become more of an issue [53].

*Collaboration and co-authorship*

In order to determine if Mileva was a collaborator or a co-author, we need to define explicitly what both of these roles entail. The Standard Collins English Dictionary defines the term collaboration as "the act of working with another or others on a joint project or something created by working jointly with another or others" [54].

However, scientific collaboration is more specific and is treated in a different manner, qualitatively it is the novel content contributed. Moreover, a collaborator and a co-author are not mutually exclusive, and so one could be classified as either or both, depending on the contribution made.

A publication from Washington University in St. Louis entitled, "Policy for authorship on scientific and scholarly publications," which was approved by the Executive Committee on Research in 2009, defined an author as "an individual who has made substantial intellectual contributions to a scientific investigation" [55]. This publication further suggests that all authors of a scientific paper should meet the following criteria: "Scholarship: Contribute significantly to the conception, design, execution, and/or analysis and interpretation of data. Authorship: Participate in drafting, reviewing, and/or revising the manuscript for intellectual content. Approval: Approve the manuscript to be published" [55].

In the context of this Washington University publication, we can see that collaboration is much broader, while co-authorship is more specific. Now we are in a position, based on all the material outlined above, to categorize Mileva's potential involvement.

## 9 Concluding Remarks

Young Einstein's early essay to his uncle Jakob in 1895 concerning the propagation of light through the ether provides insight into his deep-seated passion for physics [19]. Clearly, he had been consumed by these ideas long before going to university and also before meeting Mileva. However, Einstein's work before the Swiss Federal Polytechnic in Zürich was immature. It would take many years of study and discussion, and the intervention of Mileva in his life, before Einstein would contribute to the world of physics.

On April 10, 2004, *Physics world* ran a poll in response to the airing of the PBS Documentary, *Einstein's wife* [4]. 70% of the people surveyed believed that Mileva contributed to his work. It is impossible to say if this an indictment of the facts or merely good filmmaking [56].

Had not Mileva twice failed the Swiss Federal Polytechnic's final examination, perhaps her path would have taken a different twist. Was it Einstein that prevented Mileva from a career in physics, or was it the tertiary institution's belief that women should merely be auditors of education and not participators of it? It has also been suggested that the Swiss Federal Polytechnic examination board, which knew that Einstein and Mileva were lovers and would be married, failed Mileva on purpose, since only the husband in such a case required a degree [42]. After all, Mileva was one of the first women to undertake a theoretical physics degree in Europe in the late 19$^{th}$ century. It is naïve to believe that sexism by a stoic institution did not come into play. It is true that Marie Skłodowska-Curie gained her Ph.D. in 1903, won the Nobel Prize in Physics in the same year, and then won it again in Chemistry in 1911. She was the first person to win two Nobel Prizes, and is one of only two people in the history of the Nobels to win in two different fields, which was an incredible achievement for a woman at the beginning of the 20$^{th}$ century [57]. However, Mileva and Marie were very different people from very different backgrounds. Moreover, one person's ability cannot be measured by another's success.

In the history of physics — or even science in general — we are often confronted with paradoxes. In this instance, I am referring to Stigler's law of eponymy [58]. In its strongest and simplest form, it states, "No scientific discovery is named after its original discoverer." In fact, one could also invoke Boyer's Law, which similarly states, "Mathematical formulas and theorems are usually not named after their original discoverers" [59]. Surely in Einstein's case, this would be true. For Henry Lorentz's formulae ended up defining the special theory of relativity, and so too did Poincare's equation that Einstein redeveloped in his mass-energy equivalence paper. Boyer's Law seems to sit well with Einstein, as much as with anyone else. So why shouldn't Stigler's law of eponymy be equally plausible, concerning Mileva Marić-Einstein? Failing these suggestive statistical behaviours, perhaps we should merely invoke Occam's razor, or the law of parsimony, which states that, among competing hypotheses, the one with the fewest assumptions should be selected [60]. It seems simplest to consider that the Einsteins lived and worked together, day to day, and shared everything, including their ideas.

Albert Einstein was a genius — there is no doubt about that — but it can also be said that no man is an island. Being a genius does not mean he did not have collaborators. In fact, he collaborated later with his good friend Marcel Grossmann in order to write the general theory of relativity papers [44]. Later, he also officially worked with other scientists of the time. Mileva was a victim of her time and culture. Women belonged to their husbands and had little in the way of individual voices. It was only the very strong woman who could claim that something belonged to her, while almost all others were put in their place and simply accepted it. The oppressed usually are unaware of being such until change shows them differently.

Mileva Marić-Einstein passed away on August 4, 1948, after suffering a series of strokes that paralyzed the left part of her body. According to Milan Popović, just before the seventy-three-year-old Mileva passed away, "She kept repeating a single word, *No*" [20]. What she meant by this we will never know. However, what we do know is that much of the evidence that she collaborated with Einstein is worthy of serious consideration. A court may say that this evidence is circumstantial, but cases are won for less.

I study Einstein's work and continue to be amazed by each of his most significant papers, so much so that all my academic study is concerned with the special and general theories of relativity. I do not believe Einstein was a fraud, nor a plagiariser. In fact, I admire Einstein, and I do not judge the mistakes he made as a person, for we have all made our own. This investigation is merely an academic one to uncover a truth, and whatever it may be, I am prepared to accept it. However, I will not accept an institutionalised idea of what happened in those years, of the Einsteins, only because the mainstream deems it so.

There are so many instances that illustrate that Mileva was involved both as a collaborator and a co-author of the early papers. I would suggest that Mileva Marić-Einstein was a collaborator to all papers before the end of 1905; moreover, she was most likely a co-author of one, possibly two, significant works. The first she may have co-authored is the document of 1901, "Folgerungen aus den Kapillaritätserscheinungen" ("Conclusions Drawn from the Phenomena of Capillarity" [27]. The second paper, which is the more likely candidate, is the 1905 publication "Über einen die Erzeugung und Verwandlung des Lichtes betreffenden heuristischen Gesichtspunkt" ("On a heuristic point of view concerning the production and transformation of light") [36].

However, I cannot see that Mileva co-authored the special theory of relativity paper, as we can see a shift in Einstein's thinking and a change to a simpler mathematical form in that work. However, Mileva would have certainly been a collaborator and assistant on earlier work on special relativity, but the involvement may have been limited to mathematical treatments, not conceptual ones.

Einstein received the Nobel Prize for the photoelectric paper of 1905, which puts the recognition of that work on a very different footing. Perhaps we will never know the full extent of Mileva's involvement, but most certainly it would change the public perception of Albert Einstein. It would not lessen the genius of a great physicist, but it would provide solutions to many unanswered questions.

It would seem assigning honorary co-authorship would be an appropriate course of action in one, or perhaps two, cases, as outlined above. However, this should only be considered after an academic society or institution has made a proper and extensive inquiry and has evaluated all evidence.

Without a doubt, there was much more to Mileva Marić than we currently know, and most assuredly that is why Albert Einstein was attracted to her in the first place.